\documentclass[fleqn,twoside]{article}
\usepackage{espcrc2}

\usepackage{graphicx}
\usepackage[figuresright]{rotating}
\usepackage{epsfig}

\newcommand{\AmS}{{\protect\the\textfont2
  A\kern-.1667em\lower.5ex\hbox{M}\kern-.125emS}}

\title{ Precision simulations with  TAUOLA and PHOTOS}

\author{ Z. W\c{a}s\address[MCSD]{Institute of Nuclear Physics, Polish Academy of Sciences,\\
         ul. Radzikowskiego 152, 31-342 Cracow, Poland}%
        \thanks{  Supported in part by  the Polish Government grant KBN 1 P03 091 27   (years 2004-2006) and by
C 6th Framework Programme under contract MRTN-CT-2006-035482
(FLAVIAnet network) }}
       
\begin{document}

\begin{abstract}
The status of the Monte Carlo programs for the simulation of $\tau$-lepton production and decay in 
high-energy accelerator experiments is reviewed.
No significant changes in the organization of the programs were introduced
since previous TAU conference, that is why we will concentrate on some physical topics:
 (i) For {\tt TAUOLA} Monte Carlo generator of $\tau$-lepton decays, simulation 
of five scalar final states  based on the hadronic current became available for the 
first time. 
As an example, simple, but realistic current for final states: 
 $2\pi^-\pi^+2\pi^0\nu_\tau$, $\pi^-4\pi^0\nu_\tau$ and $3\pi^-2\pi^+\nu_\tau$
is presented. The current is installed into {\tt TAUOLA}. (ii) For the {\tt PHOTOS} Monte Carlo, which 
generates 
radiative corrections in arbitrary decays, new results  on 
next to leading order corrections became available for some decay modes.
The complete corrections  were installed for leptonic $Z$ and $B$
decays into a pair of scalars.   (iii) During conference discussions, 
the importance of 
checking the uncertainty of the overall normalization for {\tt KORALB} and {\tt  KKMC} 
simulations  was underlined. Necessary steps to check the uncertainty 
and to adjust the programs to Belle and BaBar conditions are also listed.

\vspace{2mm}
\centerline{ \it Presented at International workshop on Tau Lepton Physics, TAU06 
Pisa, Italy September,2006}
\vspace{1pc}
\centerline{preprint \hskip 1 cm  IFJPAN-IV-2006-8}
\vspace{1pc}
\end{abstract}

\maketitle

\setcounter{footnote}{0}
 
\section{Introduction}

The {\tt TAUOLA} package
\cite{Jadach:1990mz,Jezabek:1991qp,Jadach:1993hs,Golonka:2003xt}  for the simulation 
of $\tau$-lepton decays and  
{\tt PHOTOS} \cite{Barberio:1990ms,Barberio:1994qi} for the simulation of radiative corrections
in decays, are computing
projects with a rather long history. Written and maintained by 
well-defined authors, they nonetheless migrated into a wide range
of applications where they became ingredients of 
complicated simulation chains. As a consequence, a large number of
different versions are presently in use.
From the algorithmic point of view, they often
differ only in a few small details, but incorporate many specific results from distinct
$\tau$-lepton measurements. Such versions were mainly maintained (and will remain so) 
by the experiments taking precision data on $\tau$ leptons. On the other hand,
 many new applications were developed  recently,  often requiring
a program interface to other packages  (e.g. generating events for LHC, LC, 
Belle or BaBar physics processes). The programs organization,
prepared for  the convenience of users,  was presented during previous 
$\tau$ conference, in year 2004  \cite{Was:2004dg}
and we will not repeat it here. 

This time, let us concentrate on more physics oriented results.
Progress in  the simulation 
of $\tau$ decays into final states of five scalars  
was achieved \cite{Kuhn:2006nw} recently. This work has some consequences 
for the general form how the Monte Carlo programs for decay chains have to be 
organized. We will devote section 2 to discussion of that subject.
In section 3 we will present some new results for the simulation 
of radiative corrections in decays with {\tt PHOTOS} Monte Carlo. Here papers 
\cite{Nanava:2006vv,Golonka:2006tw}, where Next to Leading Order effects 
were introduced into generation for the first time, will be presented.
 Even though those results do not affect simulation of 
$\tau$ decays in the direct way, nonetheless open the way for future
precision statements on radiative effects in $\tau$ decays.
That is why they should find their place in $\tau$ conference proceedings.

During my talk and in discussions later, I found significant interest 
in domain of overall normalization of $\tau$-lepton pair 
production cross sections as calculated by {\tt KORALB} \cite{Jadach:1984iy,Jadach:1985ac}  and {\tt KKMC} \cite{Jadach:1999vf} generators.
These programs were written for higher energies. For the use at
Belle, BaBar energies some effort necessary to adjust photon vacuum polarization 
 need to be completed. Section 3 is devoted to that purpose.
List of necessary benchmarks is  given and necessary steps 
are explained.

Because of the limited space of the contribution, 
and sizable amount of other physically interesting 
results, some of them will be excluded from conference 
proceedings.In particular, this time we will skip completely 
the applications related to use of $\tau$ physics in domain of searches for new physics.
\section{Five pion final states in  {\tt TAUOLA} Monte Carlo}
Approximations were used for five scalar  decay modes 
 in {\tt TAUOLA} until recently.
In relative recent preprint \cite{Kuhn:2006nw}, the structure for implementation of 
hadronic currents into $\tau$ decay modes of five scalars was prepared. 
For the new code,
technical
tests of the solution including benchmark distribution were performed and described. 
Finally,
currents were prepared for some decay modes.
For the $2\pi^-\pi^+2\pi^0$ mode the three decay chains
$\tau^- \to a_1^- \nu \to 
\rho^-(\to \pi^-\pi^0) \omega (\to \pi^-\pi^+\pi^0)  \nu$,
$\tau^- \to a_1^- \nu \to 
a_1^-(\to 2\pi^-\pi^+) f_0 (\to 2\pi^0)  \nu$, and
$\tau^- \to a_1^- \nu \to 
a_1^-(\to \pi^-2\pi^0) f_0 (\to \pi^+\pi-)  \nu$ are introduced
with simple assumptions about the couplings and propagators of the
various resonances. Similar amplitudes (without the
$\rho\omega$ contributions) are adopted for the $\pi^-4\pi^0$
and $3\pi^-2\pi^+$ modes. 

The five-pion amplitude is thus based on a simple model, 
which, however, can be considered as a first realistic  example.
As usual, hadronic current
is easy to replace by the more sophisticated one, once it is required. 
Also, multitude of additional decay channels of five scalars (pions or K-ons) are 
pre-installed. Appropriate flavours and  hadronic currents have to be 
provided by the user.

From that perspective, and in general from the point of view of any future Monte Carlo 
program, it is of some interest to show some of the numerical results. We will
concentrate mainly on effects of the different types of interferences.

{ 
\begin{figure}
\begin{center}
\begin{tabular}{|r|r|r|r|r|}
\hline
   & Final state  & Current &{\scriptsize $\Gamma_X/\Gamma_e\times 10^3$} 
                                   & {\scriptsize $\Gamma_X/\Gamma_e \times 10^3$ }
                                   \\
          &              &         & {\scriptsize \tt (TAUOLA) }   &
                                   (exp. )
                                                                    \\ 
\hline
1& $2\pi^-\pi^+2\pi^0$ &A  no s. &24.04  & --  \\
2& $2\pi^-\pi^+2\pi^0$ &B  no s. &  ${ \bullet} \;$ 9.28  & -- \\
\hline
3& $2\pi^-\pi^+2\pi^0$ &A  s.  &25.30  & $25\pm3$   \\
4& $2\pi^-\pi^+2\pi^0$ &B   s. & 6.05  & $6.2\pm2$   \\
5& $2\pi^-\pi^+2\pi^0$ &A+B  s. &31.35  & $31\pm2$            \\
\hline 
6&   $\pi^-4\pi^0$     &B  s. & 9.37  & $5.5^{+3.4}_{-2.8}$
                                               \\
7& $3\pi^-2\pi^+$      &B  s.  & 11.03  & $4.6\pm0.3$   \\
\hline
\end{tabular}\vspace{0.3cm}
\end{center}
Table 1: {\it   Test results of the generator for realistic choices of
parameters; see the text of the paper \cite{Kuhn:2006nw} for details. 
Two currents {\bf A} and  {\bf  B } are used in different combinations.
The s. (or no s.) comment, denotes that for the particular channel symmetrization over 
identical pions was (was not) included. With the $\bullet$ we denote the result
difficult to interpret, because of   $m_{\pi^0} \ne m_{\pi^\pm}$   see the text.
}
\end{figure}}

In the numerical results, collected in Table 1 the following two currents are used.
{\bf  Current A:} $\tau^- \to a_1^- \nu \to 
\rho^-(\to \pi^-\pi^0) \omega (\to \pi^-\pi^+\pi^0)  \nu$, 
{\bf Current B:} $\tau^- \to a_1^- \nu \to 
a_1^-(\to \pi^-, \pi^-\pi^+ (2\pi^0) ) f_0 \bigl(\to 2\pi^0 (\pi^+\pi^-)\bigr)  \nu$. 
If we  compare lines 3, 4 and 5 of the Table, we can see that the 
interference between current $A$ and $B$ seems to be zero (or very small) on total rate.
It is nice and encouraging result. This 
picture is different
when one compare lines 1 with 3,  or 2 with 4. There, interference
which appears  due to symmetrization over identical pions in final state
is respectively +5 \% and -11 \%. In case of current B, data from the Table 
are not easy to interpret, because of statistical factor $\frac{2}{3}$ and 
 $m_{\pi^0} \ne m_{\pi^\pm}$. What is more, the  symmetrization effect of  
-11 \%  is obscured by 
$m_{\pi^0} \ne m_{\pi^\pm}$ effect. 

Let us turn now to more general discussion of 
our  numerical results which can  be compared with the experimental data of ref. 
\cite{Eidelman:2004wy}. 
Perfect agreement in case of  lines 3, 4, and 5 is a trivial consequence 
of the fact that these
data points were used as our input for the current parameters. On the other hand,
modest agreement of {\tt TAUOLA} with the data  (lines 6 and 7), provide
 test of the predictive power of our model.

Rather unpredictive nature of interference effects is of importance in 
construction of Monte Carlo programs, if they are supposed 
 to be used in comparisons with the precision data 
on decays. The constructive/destructive interferences originate from the 
corrections of order  $\Gamma/Q$ for intermediate resonances, which  appear 
in the cascade 
decays. If widths of the resonances saturating currents would be sufficiently small
then of course interferences would be negligibly small. At the same time the intermediate
states would be well formed and necessities to use different parametrizations for example 
of $\rho$ resonance, depending on how it was formed and how it will decay, would not 
be necessary. 

Unfortunately, in practice, it is not the case, and we have to bear this 
 constraint unpleasant
 for program construction in mind.

\section{{\tt PHOTOS} and NLO effects in $B$ and $Z$ decays}
\def\CCol{{\tt SANC}}
There were significant changes introduced into  {\tt PHOTOS} Monte Carlo project over the last
two years. The complexity of the subject matches neither size nor the  purpose 
of the present talk. Recently collected results \cite{Nanava:2006vv,Golonka:2006tw}
only indirectly affect predictions of simulation for $\tau$ physics. However as for the first time explicit form of the approximate matrix elements actually used was 
written, new possibilities
opened. In particular,   for the two-body decay modes into
 fermions ($Z\to \mu^+\mu^- $) or scalars (eg. $B\to K^+\pi^- $) exact NLO
order matrix elements were implemented into {\tt PHOTOS} kernel. In cases when
 the correction kernels were switched on and simulations were restricted to the 
first order,
the differences  with results of reference Monte Carlo programs were below statistical
errors of $10^8$ event samples. Also technical tests of {\tt PHOTOS} performed at that or even better
statistical level confirmed the correctness of program design; 
even though significant changes in crude distributions were introduced, the results 
did not change at all.

To visualize the results let us present the example plots for  ($Z\to \mu^+\mu^- $)
and  $B\to K^+\pi^- $ cases.
In case of ($Z\to \mu^+\mu^- $) we could perform comparisons for multi-photon version
of the generation, because our reference program {\tt KKMC} has such possibility as well.
 The histogram with the worst agreement of all possible 
invariant masses which could be constructed from four momenta of: $\mu^+, \mu^- $
and eventually up to two hardest photons (their energies had to be above 1 GeV)
is shown in Fig 1 where the invariant mass of $\gamma\gamma$ pair is plotted.
Even in this case, the green and red  (gray) lines for {\tt PHOTOS} and {\tt KKMC}  
nearly overlap (the logarithmic left side scale has to be used for that lines).
To visualize the differences, we show on the plot the ratio of the two histograms
which is depicted with the black line and follows  linear right scale.
The differences in the ratio are rather small for the masses of $\gamma\gamma$ pair 
up to about 50 GeV. For larger masses which contribute about $10^{-3}$ to the sample 
of two-photon configurations the difference gradually grow to about 15\% close to
the phase space limit. Note that
fraction of  events with  least two photons each of which with energies above 1 GeV is
1.26 \% from {\tt KKMC}  and 1.29 \% from {\tt PHOTOS}. 

In case of the $B$ meson decays the agreement between {\tt PHOTOS} and reference calculation,
even without the use of correcting weight was excellent, that is better than for the presented
above result for $Z$ decay. That is why we skip numerical result and address the reader to 
conference transparencies or to ref. \cite{Nanava:2006vv}.

\begin{figure}
\begin{center}
\setlength{\unitlength}{0.5 mm}
\begin{picture}(35,80)
\put( -65,-45){\makebox(0,0)[lb]{\epsfig{file=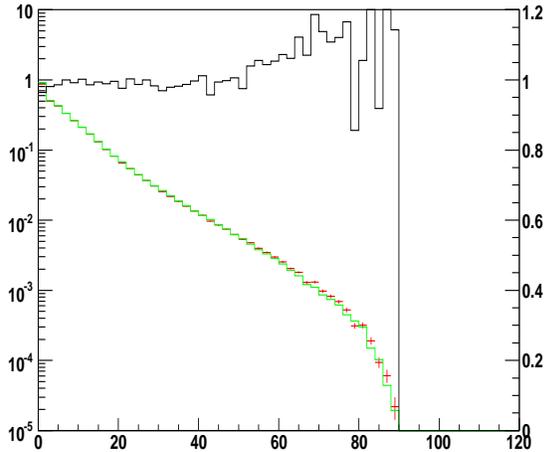,width=80mm,height=70mm}}}
\end{picture}
\end{center}
\vskip 1.5 cm
\caption{\small \it   Comparisons of improved  {\tt PHOTOS} with multiple photon emission and {\tt KKMC}  with second order
matrix element and exponentiation.  Comparison figure of worst agreement  
was  selected from all constructed with the help of {\tt MC-TESTER} \cite{Golonka:2002rz} for details 
see the text of the paper.
 } 
\end{figure}
\section{ { Normalization issues for Monte Carlo programs for $\tau$-lepton pair production at Belle/BaBar energies}}
In discussions 
it became obvious that some comments on the overall normalization 
of  {\tt KORALB} and {\tt KKMC} predictions  
for $\tau$-lepton pair production at Belle and BaBar energies is of importance
for the present day users.

One has to bear in mind that {\tt KORALB} was published \cite{Jadach:1984iy,Jadach:1985ac} more than {\it twenty years ago}. The program was supposed to feature
orthodox first order QED corrections and complete  mass and spin 
effects. Such formula turned out to be very useful, and program remains in 
broad use until now. On the other hand, some of its inputs are rather outdated
and do not match the present day requirements, even for technical tests.
To be precise, I have in mind the function {\tt PIRET(S)}, which features the
real part of hadronic vacuum polarization of photon as measured by the data
collected until early 80's. Obviously this function need to be replaced
by the more modern one. A possible choice can be the function
{\tt REPI} of ref. \cite{Burkhardt:2005se} because it 
t has similar functionality as
{\tt PIRET} of {\tt KORALB}.

Unfortunately the  improvements on  {\tt FORTRAN} function {\tt PIRET(S)}
describing hadronic vacuum polarization of photon, 
 do not solve all
normalization problems of {\tt KORALB}. It is well known since long,
that the genuine one loop corrections are not enough, and the solutions
are available. The two major improvements which appeared as a consequence 
of phenomenology improvements during the LEP era was introduction of higher
order QED corrections into Monte Carlo simulations and better way
to combine loop corrections with the rest of the field theory calculations.
It was found to be safe  to sum contributions of loop 
corrections into photon (and $Z$) propagators. Then, 
terms of all, but incomplete, orders of perturbation expansion
are taken into account. That is 
why
significant effort was needed to justify the approach \cite{Bardin:1999ak}.
 At lower energies things are of course simpler, as there is no need to worry 
about $Z$-lineshape. The {\tt KKMC} Monte Carlo \cite{Jadach:1999vf} 
could  thus be a complete solution
to B-factory needs and ready to use. This program features higher order QED
matrix elements with the help of exclusive exponentiation, and explicit matrix 
element up to the second order. Unfortunately electroweak library
(which include vacuum polarization, the only function interesting for low
 energy applications) was never adopted to requirements of processes
below 13 GeV center-of-mass energies. 

To provide reasonably good results, this library needs thus to be overruled
(analogously to the function {\tt PIRET(S)} of {\tt KORALB}) by a more 
suitable 
one\footnote{
In {\tt KKMC} similar improvements require to  overwrite
in routine
{\tt DZface{\_}MakeGSW} (file {\tt dizet/Dzface.f})
calculation of {\tt GSW(6) }
with the one using function {\tt PIRET(SS)} or {\tt REPI(SS)}. The  {\tt PIRET(SS)} must 
be supplemented with the leptonic contribution to vaccum polarization as well. 
It was checked by S. Banerjee  that pretabulation (in contrast of low energy
numerical 
values of KKMC vaccum polarization)  is not a problem. To 
this end he simply increased 
density of pretabulation points.
}.  
After small tests the function {\tt REPI} of ref. \cite{Burkhardt:2005se} 
should be suitable
(overall normalization constants and other details of conventions
need to be checked).
 
Once this is completed, and if 
two-loop photon vacuum polarization can be neglected, 
{\tt KORALB} and {\tt KKMC} can form a base for tests and studies of systematic errors for cross section normalizations 
at low energies. The necessary strategy should be 
 similar to the one for the Bhabha scattering see eg. \cite{Jadach:1991pj}
for description.

For that purpose 
the following calculations
of the total cross section  should be performed: 
\begin{enumerate}
\item
{\tt KORALB}  radiative 
corrections switched off. 
\item {\tt KKMC}, both  
electroweak corrections and initial state QED bremsstrahlung switched off
(final state bremsstrahlung need to be kept on, 
otherwise spin amplitudes necessary for calculation of $\tau$ spin correlations are
not calculated and program stops). 
\item 
Radiative corrections switched on in {\tt KORALB}.
\item
Radiative corrections  switched on in {\tt KORALB}.
Vacuum polarization  switched off with the help of internal 
key {\tt IFVAC=0}. 
\item
{\tt KKMC}:  
electroweak corrections on,   initial state QED bremsstrahlung switched off.
 \item
{\tt KKMC}:  
electroweak corrections switched off but initial state QED bremsstrahlung switched on
\item
{\tt KKMC}: both  
electroweak corrections and initial state QED bremsstrahlung switched on
\end{enumerate}

The results, we will call them respectively $\sigma_1 ...\sigma_7 $,
can be  calculated for $\tau$- or $\mu$-pair production. Also experimental
cuts can be applied in the calculation of these cross section. 
Let us list now some of the possible checks. Of course all calculations
being compared 
must be performed with the same assumptions on experimental cuts and final 
state flavours.

$\bullet$: $\sigma_2$ should be larger by a factor of 
$1+\frac{3}{4}\frac{\alpha}{\pi}$ than $\sigma_1$ for all center of mass energies in case 
no experimental cuts are applied. $\bullet$: the following
 relation\footnote{In case of {\tt KORALB} vacuum polarization
increase the cross section by   $(2\Re \Pi_{\gamma,\gamma}(s))$
times Born cross section.  In case of {\tt KKMC} the 
vacuum polarization factor $\frac{1}{(1- \Pi_{\gamma,\gamma}(s'))^2}$ appears
for all events. In case when initial state bremsstrahlung is switched on,
the effective transfer $s'$ depends
on the amount of energy carried out by the initial state radiation 
convolution with  initial state radiation spectrum is formed. } 
should hold:
$\sigma_3- \sigma_4=\sigma_1 2 \bigl(1- \sqrt{\frac{\sigma_2}{\sigma_5}}\bigr)$.
 $\bullet$:
finally comparisons of $\sigma_6$  $\sigma_4$ with $\sigma_1$ can be used 
to estimate the size of QED bremsstrahlung effects respectively of higher orders 
or just first order alone.
$\bullet$: the  relation  $\frac{\sigma_7}{\sigma_5}=\frac{\sigma_6}{\sigma_2}$
is not expected to hold precisely.
It can hint on numerical  importance of convoluting  QED bremsstrahlung
and vacuum polarization with respect to naive factorization. 

If comparisons are repeated with 
experimental  cuts  applied, some extra care must be taken, because of cut off dependence of  final state bremsstrahlung effects.
On the other hand, numerically significant and theoretically 
ambiguous  contributions from events with
very low final state lepton pair invariant masses, are removed.

Unfortunately the above points must remain in sketchy form. 
Full clarification, 
as LEP experience showed, require significant amount of work.
 
\section{Summary and future possibilities}

The status of the computer programs for the decay of $\tau$ leptons and
associated projects  was reviewed.
The high-precision version of  {\tt PHOTOS} for radiative corrections was presented.
 In particular, the option to run the program with multiple-photon radiation
was mentioned. 
New results for  leptonic decays of $Z$, and $B$ meson decays into pair 
of scalars, were presented.  For these channels complete next-to-leading order
effects can now be  simulated. However, for the most of the applications these effects
are not necessary, leaving standard modular version of {\tt PHOTOS} sufficient.
 The important result of the above work, is that the path to include 
electromagnetic form-factors of the particles participating in decay 
 is now open 
for the future fits to the data. These form-factor effects may be significantly larger 
and physically more justified, than  complete next-to-leading order
effects of scalar QED in $B$ meson decays recently installed,,

 The presentation of the {\tt TAUOLA} general-purpose interface
  was omitted because of lack of time. Examples for its use in the case of the Higgs boson parity
measurement at a future Linear Collider \cite{Bower:2002zx,Desch:2003mw,Desch:2003rw} 
and for Higgs searches at the LHC \cite{Richter-Was:2004jf} can be found 
in the literature.  Recently, a similar application was  developed for the
case of studies in hypothetical effects of  CP-parity breaking in the $B_0$-$\bar B_0$ system
at Belle and BaBar \cite{Chankowski:2004tb}.

Distinct versions of the {\tt TAUOLA} library for $\tau$ lepton decay, and of
{\tt PHOTOS} for radiative corrections in decays, are now in use.
 The principles how to use  the distribution
package, are presented in refs. \cite{Golonka:2003xt,Was:2004dg}.

In the  talk we have reviewed the results for the novel decay modes 
of {\tt TAUOLA} into
five scalars. These modes  feature now the hadronic current. The simple  
but realistic current \cite{Kuhn:2006nw} is available for decay modes
 $2\pi^-\pi^+2\pi^0\nu_\tau$, $\pi^-4\pi^0\nu_\tau$ and $3\pi^-2\pi^+\nu_\tau$.
Numerical study of the new decay modes helped to formulate
 comments on the importance of $\Gamma/M$ terms for the intermediate resonances.
It is argued that the  constructive or destructive interferences 
appear and that   their existence must be taken into account by builders 
of the future 
Monte Carlo programs for decays (not necessarily $\tau$ decays).

Finally, presentation of adjustments for {\tt KORALB} and {\tt KKMC} programs 
in treatment of photonic vacuum polarization was given. It was explained,
 that without such changes the programs can not be used 
for discussion  of the normalization uncertainty for  the $\tau$-lepton pair production
cross section  at Belle/BaBar energies.
\vskip 2 mm
\centerline{ \bf Acknowledgements}
\vskip 1 mm
 Useful collaboration  and suggestions from 
 E. Barberio, G. Nanava, S. Jadach, J. H. K\"uhn, S. Eidelman, B. Kersevan,  E. Richter-Was
P. Golonka are acknowledged. Discussions   
with  members of the Belle and BaBar collaborations 
are also acknowledged. Finally, exchange of e-mails with  B. Pietrzyk, H. Burkhardt,
D. Bardin, T. Riemann, S. Banerjee and M. Roney was a necessary input to section 4 of the present 
report.

\providecommand{\href}[2]{#2}


\end{document}